\documentclass[12pt]{article}
\usepackage{amsfonts}

\usepackage{amssymb}
\usepackage{amsmath}
\usepackage{amsthm}
\usepackage{textcomp}
\usepackage{ccfonts}
\usepackage{psfrag}
\usepackage{epsfig}
\usepackage{tikz,color}
\usepackage{latexsym}
\usepackage{amscd}

\newtheorem{prop}{Proposition}

\newtheorem{theorem}{Theorem}

\def\PROB {{\mathbb P}}
\def\EXP {{\mathbb E}}
\def\IND{{\mathbb I}}
\def\Var{{\mathbb Var}}

\def\A{{\mathbf{A}}}
\def\bB{{\mathbf{B}}}

\def\t{\intercal}

\def\D{{\cal D}}

\def\R{{\mathbb R}}
\def\Rd{{\mathbb R}^d}

\def\bX{{\mathbf{X}}}
\def\b1{{\mathbf{1}}}
\def\bx{{\mathbf{x}}}
\def\bz{{\mathbf{z}}}
\def\ba{{\mathbf{a}}}
\def\bb{{\mathbf{b}}}
\def\bc{{\mathbf{c}}}
\def\bC{{\mathbf{C}}}
\def\be{{\mathbf{e}}}
\def\bE{{\mathbf{E}}}
\def\bL{{\mathbf{L}}}

\def\bQ{{\mathbf{Q}}}
\def\bS{{\mathbf{S}}}

\def\bZ{{\mathbf{Z}}}
\def\bV{{\mathbf{V}}}

\def\argmin{\mathop{\rm arg\, min}}
\def\argmax{\mathop{\rm arg\, max}}

\begin{document}
\begin{titlepage}
\thispagestyle{empty}
\setcounter{page}{0}

\title{Repeated observations for classification}
\author{Huseyin Afser\thanks{Adana Alparslan T\"urkey Science and Technology University, afser@atu.edu.tr}
\and L\'aszl\'o Gy\"orfi\thanks{Budapest University of Technology and Economics, gyorfi@cs.bme.hu.
The research of L\'aszl\'o Gy\"orfi has been supported by the National Research, Development and Innovation Fund of Hungary  under the 2020-1.1.2-PIACI-KFI funding scheme.}
\and Harro Walk\thanks{Universit\"at Stuttgart, harro.walk@mathematik.uni-stuttgart.de}}
\maketitle

\begin{abstract}
We study the problem nonparametric classification with repeated observations.
Let $\bX$ be the $d$ dimensional feature vector and let $Y$ denote the  label taking values in $\{1,\dots ,M\}$.
In contrast to usual setup with large sample size $n$ and relatively low dimension $d$, this paper deals with the situation, when
instead of observing a single feature vector $\bX$ we are given $t$ repeated feature vectors $\bV_1,\dots ,\bV_t $.
Some simple classification rules are presented such that the conditional error probabilities have exponential convergence rate of convergence as $t\to\infty$.
In the analysis, we investigate particular models like robust detection by nominal densities, prototype classification, linear transformation, linear classification, scaling.
\end{abstract}

\noindent


\end{titlepage}

\section{Introduction}
\label{Intr}

As in the basic setup of classification, let the feature vector
$\bX$ take values in $\Rd$, and let its label
$Y$ take values $1,\dots , M$.
The task of classification is to decide
on $Y$ given $\bX$, i.e., one aims to find a decision function $g$ defined on the range of $\bX$ such
that $g(\bX)=Y$ with large probability.
If $g$ is an arbitrary decision function
then its error probability is denoted by
\[
P_e(g)=\PROB\{g(\bX)\ne Y\}.
\]

It is well-known that
the Bayes decision $g^*$ minimizes the error probability:
\[
g^*(\bx) =\argmax_j \PROB\{Y=j\mid \bX=\bx\}
\]
and
\[
P_e^*=\PROB\{g^*(\bX)\ne Y\}=\min_g P_e(g)
\]
denotes its error probability.
(In case there are several indices achieving the
maximum, choose the smallest one.)

The Bayes decision cannot be constructed as long as the distribution
of $(\bX,Y)$ is unknown. Assume, that we observed data
\begin{align}
\label{data}
\D_n=\{(\bX_1,Y_1), \dots ,(\bX_n,Y_n)\}
\end{align}
consisting of independent and identically distributed copies of $(\bX,Y)$.
Devroye, Gy\"orfi and Lugosi
\cite{DeGyLu96} contains classification algorithms with universal
consistency properties, which means that the error probability of these algorithms tends to the Bayes error probability for all distribution of $(\bX,Y)$.

In contrast to the usual theory, where the sample size $n$ is much larger than the (small) dimension $d$, here we consider the case of large $d$.
Thus, one cannot aim to approach the Bayes error probability.
In this note we study the possible decrease of the error probability for having $t$ repeated feature vectors.
For sensors, it means that each sensor  generates $t$ samples, i.e., the classification is based on multidimensional time series.
\textcolor{black}{In this paper we present simple classification rules, the error probability of which tends to zero exponentially fast as $t\to \infty$,
i.e., for the error probability, we derive exponential upper bounds of form $e^{-E_dt}$ with error exponent $E_d>0$.
Interestingly, the error exponent might be increasing with $d$.}

As in Horv\'ath et al. \cite{HoKoMoNo20}, a motivating example can be the modeling of sensor networks, containing $d$ sensors of various type: radar, optic, acoustic, motion, bio, chemical, etc.
A challenging feature of these problems is that the conditional distributions of the feature vector $\bX$ may change from time to time or the actual set of sensors depends on the meteorological conditions.

The paper is organized as follows.
In Section 2 we introduce two classifiers with exponential rate of convergence of the error probability.
As  illustrations, Section 3 is on classifier classification derived from nominal densities, while in Section 4 we analyse a prototype classification.
Finally, in Section 5 we study the consequences for linear classification.

\section{Classification for repeated observations}
\label{repeat}

\textcolor{black}{For the two hypotheses} testing and in case of $t$ repeated observations, Stein and Chernoff constructed tests such that both errors, the error of the first and second kind tend to zero exponentially fast as $t\to \infty$, see Chernoff \cite{Che52} and Chapter 11 in Cover and Thomas \cite{CoTh91}.
These results can be extended to multiple hypotheses such that
the $d$ dimensional feature vector $\bX$ is replaced by the $d$ dimensional observation vectors $\bV_1,\dots ,\bV_t$, and the decision on $Y$ is based on $\bV_1,\dots ,\bV_t$. For sensor networks, this setup is interpreted as $t$ multivariate samples.

Let $\nu_j$ be the conditional distribution of $\bX$ given $Y=j$:
\begin{align*}
\nu_j(A)=\PROB\{\bX\in A\mid Y=j\}.
\end{align*}
If $f_j$ stands for the density of $\nu_j$ with respect to a dominating measure $\lambda$, then the maximum likelihood (ML) decision is defined by
\begin{align*}
g_{ML}(\bV_1,\dots ,\bV_t)
&=
\argmax_j \prod_{i=1}^t f_j(\bV_i).
\end{align*}

If for given $Y=j$,   $\bV_1,\dots ,\bV_t $ are conditionally independent and identically distributed, $j=1,\dots , M$, then let's bound the error probability of the ML decision.
The definition of the ML decision implies that
\begin{align*}
\PROB\{ g_{ML}(\bV_1,\dots ,\bV_t) \ne Y\}
&\le
\max_{j }\PROB\{ g_{ML}(\bV_1,\dots ,\bV_t)\ne j\mid Y=j\}\\
&=
\max_{j}\sum_{\ell\ne j }\PROB\{ g_{ML}(\bV_1,\dots ,\bV_t)=\ell\mid Y=j\}\\
&=
(M-1) \max_{j} \max_{\ell\ne j }\PROB\left\{\prod_{i=1}^t f_{\ell}(\bV_i)>\prod_{i=1}^t f_j(\bV_i) \mid Y=j\right\}.
\end{align*}
Apply the Chernoff's bounding technique:
\begin{align*}
\PROB\left\{\prod_{i=1}^t f_{\ell}(\bV_i)>\prod_{i=1}^t f_j(\bV_i) \mid Y=j\right\}
&\le
\EXP\left\{\sqrt{\frac{\prod_{i=1}^t f_{\ell}(\bV_i)}{\prod_{i=1}^t f_j(\bV_i)}} \mid Y=j\right\}\\
&=
\prod_{i=1}^t\EXP\left\{\sqrt{\frac{ f_{\ell}(\bV_i)}{ f_j(\bV_i)} }\mid Y=j\right\}\\
&=
\EXP\left\{\sqrt{\frac{ f_{\ell}(\bV_1)}{ f_j(\bV_1)}} \mid Y=j\right\}^t\\
&=
\left[\int \sqrt{\frac{ f_{\ell}(\bx)}{ f_j(\bx)}} f_j(\bx)\lambda(d\bx) \right]^t \\
&=
\left[\int  \sqrt{f_{\ell}(\bx) f_j(\bx)} \lambda(d\bx) \right]^t.
\end{align*}
\textcolor{black}{This yields a bound on the error exponent:
\begin{align}
-\frac{1}{t} \log \left( \PROB\{ g_{ML}(\bV_1,\dots ,\bV_t) \ne Y\} \right) \geq    \min_{{\ell} \neq j} \mathbb{B}(f_j,f_{\ell})  - \frac{1}{t} \log(M-1),
\end{align}}
where
\begin{align*}
\mathbb{B}(f_j,f_{\ell}) = - \log \int  \sqrt{ f_{\ell}(\bx) f_j(\bx) } \lambda(d\bx)
\end{align*}
is the Bhattacharyya distance between densities $ f_{\ell}$ and $ f_{j}$.

In the standard setup of classification these densities are unknown.
Afser \cite{Afs2021} formulated the problem of classification for repeated observations.
A classifier $g$ is called elementary, if it  operates on $\bX$, i.e., the corresponding classification rule has the form $g(\bX)$.
For an elementary classifier $g$, introduce the aggregated classifier $\tilde g$ as follows:
\begin{align*}
\tilde g(\bV_1,\dots ,\bV_t)
&=
\argmax_j \sum_{i=1}^t \IND_{g(\bV_i)=j},
\end{align*}
where $\IND$ denotes the indicator function.
Thus, $\tilde g$ operates on the repeated observation $\bV_1,\dots ,\bV_t$.
Next we show that if the elementary classifier $g$ has nontrivial conditional error probabilities, then the conditional error probabilities of the aggregated classifier $\tilde g$ tend to zero exponentially fast as $t\to \infty$.

\begin{theorem}
\label{multi}
Assume that for given $Y=j$,   $\bV_1,\dots ,\bV_t $ are conditionally independent and identically distributed, $j=1,\dots , M$.
Put
\begin{align*}
p_{j,\ell}=\PROB\{g(\bV_1)=\ell\mid Y=j\}=\nu_j(\{\bx:g(\bx)=\ell\}).
\end{align*}
If $p_{j,j}> p_{j,\ell}$ for all $j\ne \ell$,  then
\begin{align*}
\PROB\{\tilde g(\bV_1,\dots ,\bV_t) \ne Y\}
&\le
\max_{j }\sum_{\ell\ne j} \left( 1-(\sqrt{p_{j,j}}-\sqrt{p_{j,\ell}})^2\right)^t.
\end{align*}
\end{theorem}

\begin{proof}
One has that
\begin{align*}
\PROB\{\tilde g(\bV_1,\dots ,\bV_t) \ne Y\}
&\le
\max_{j }\PROB\{\tilde g(\bV_1,\dots ,\bV_t)\ne j\mid Y=j\}\\
&=
\max_{j}\sum_{\ell\ne j }\PROB\{\tilde g(\bV_1,\dots ,\bV_t)=\ell\mid Y=j\}.
\end{align*}
Furthermore,
\begin{align*}
\PROB\{\tilde g(\bV_1,\dots ,\bV_t)=\ell\mid Y=j\}
&\le
\PROB\left\{\sum_{i=1}^t \IND_{g(\bV_i)=\ell}\ge \sum_{i=1}^t \IND_{g(\bV_i)=j}\mid Y=j\right\}\\
&=
\PROB\Big\{\sum_{i=1}^t \left(\IND_{g(\bV_i)=\ell}- \IND_{g(\bV_i)=j}\right)\ge 0\mid Y=j\Big\}.
\end{align*}
For $\beta>0$, the Chernoff type large deviation technique implies that
\begin{align*}
&\PROB\{\tilde g(\bV_1,\dots ,\bV_t)=\ell\mid Y=j\}\\
&\le
\EXP\left\{e^{\beta\sum_{i=1}^t \left(\IND_{g(\bV_i)=\ell}- \IND_{g(\bV_i)=j}\right)}\mid Y=j\right\}\\
&=
\EXP\left\{e^{\beta \left(\IND_{g(\bV_1)=\ell}- \IND_{g(\bV_1)=j}\right)}\mid Y=j\right\}^t\\
&=
\Big(e^{\beta}\PROB\{g(\bV_1)=\ell\mid Y=j\}\\
&\quad + e^{-\beta}\PROB\{g(\bV_1)=j\mid Y=j\}+\PROB\{g(\bV_1)\ne\ell, g(\bV_1)\ne j\mid Y=j\}\Big)^t.
\end{align*}
Thus,
\begin{align*}
\PROB\{\tilde g(\bV_1,\dots ,\bV_t)=\ell\mid Y=j\}
&\le
\left(e^{\beta}p_{j,\ell}+ e^{-\beta}p_{j,j}+1-p_{j,\ell}-p_{j,j}\right)^t.
\end{align*}
For the choice $e^{\beta}=\sqrt{\frac{p_{j,j}}{p_{j,\ell}}}>1$ we get that
\begin{align*}
\PROB\{\tilde g(\bV_1,\dots ,\bV_t)=\ell\mid Y=j\}
&\le
\left(2\sqrt{p_{j,j}p_{j,\ell}}+1-p_{j,\ell}-p_{j,j}\right)^t
=
 \left( 1-(\sqrt{p_{j,j}}-\sqrt{p_{j,\ell}})^2\right)^t.
\end{align*}
\end{proof}

If $\PROB\{g(\bV_1) \ne Y\}=0$, then $p_{j,j}=1$ and $p_{j,\ell}=0$, for all $j\ne \ell$, and therefore the bound in the theorem is $0$, too.

The large deviation inequality in the proof is tight such that Sanov's theorem implies that
\begin{align*}
\frac 1t \ln\PROB\{\tilde g(\bV_1,\dots ,\bV_t)=\ell\mid Y=j\}
&\to
\ln\left( 1-(\sqrt{p_{j,j}}-\sqrt{p_{j,\ell}})^2\right),
\end{align*}
therefore
\begin{align*}
\lim_t \frac 1t \ln\PROB\{\tilde g(\bV_1,\dots ,\bV_t)\ne Y\}
&=
\max_{j\ne \ell}\ln\left( 1-(\sqrt{p_{j,j}}-\sqrt{p_{j,\ell}})^2\right).
\end{align*}
There are some refinements of this asymptotics.
For example,
\begin{align*}
\PROB\{\tilde g(\bV_1,\dots ,\bV_t)=\ell\mid Y=j\}
&=
\frac{\left(1-(\sqrt{p_{j,j}}-\sqrt{p_{j,\ell}})^2 \right)^t }{\left(  {2\pi} t  \right)  ^{1/2}c_{\ell,j}}%
\left(1+O\left(  \frac{1}{t}\right)\right)
\end{align*}
with constant $c_{\ell,j}$ depending on the probabilities $p_{j,\ell}$ and $p_{j,j}$, see Bahadur and Rao \cite{BaRa60}.

If for a classifier $g$,
\begin{align*}
\min_{j }\PROB\{g(\bV_1)=j\mid Y=j\}>1/2,
\end{align*}
then by simplifying the proof of the theorem one gets a slightly worse bound:
\begin{align}
\PROB\{\tilde g(\bV_1,\dots ,\bV_t) \ne Y\}
&\le
\max_{j }\PROB\{\tilde g(\bV_1,\dots ,\bV_t)\ne j\mid Y=j\}\nonumber\\
&\le
\max_{j }\PROB\Big\{\sum_{i=1}^t \left(\IND_{g(\bV_i)\ne j}- \IND_{g(\bV_i)=j}\right)\ge 0\mid Y=j\Big\}\nonumber\\
&\le
\max_{j } \left(2\sqrt{p_{j,j}(1-p_{j,j})}\right)^t.
\label{M}
\end{align}

For some applications, instead of a strict bound an approximation is more useful:
\begin{prop}
\label{CLT1}
Under the conditions of Theorem \ref{multi}, the Central Limit Theorem (CLT) implies that
\begin{align*}
\PROB\{\tilde g(\bV_1,\dots ,\bV_t)=\ell\mid Y=j\}
&\le
\PROB\Big\{\sum_{i=1}^t \left(\IND_{g(\bV_i)=\ell}- \IND_{g(\bV_i)=j}\right)\ge 0\mid Y=j\Big\}\\
&\approx
\Phi\left(-\frac{\sqrt{t}(p_{j,j}-p_{j,\ell})}{\sqrt{\Var\left(\IND_{g(\bV_1)=\ell}- \IND_{g(\bV_1)=j}\mid Y=j\right)}}\right),
\end{align*}
where $\Phi$ stands for the standard normal distribution function. Furthermore, the Berry-Esseen inequality yields that the approximation error is of order $O(1/\sqrt{t})$.
\end{prop}

\bigskip
A generalization of the above ML/log-likelihood decision functions and the elementary classifier is given by elementary utility functions $h_j : \mathbf{R}^d \rightarrow \mathbf{R}$, $j=1,\cdots,M$, where $h_j$ measures the statistician's trust in the label $j$ on the basis of an observation $\bx$. The aggregated decision function $\widehat g$ is  introduced by
\begin{align*}
\widehat g (\bV_1,\cdots,\bV_t ) = \argmax_{j} \sum_{i=1}^t h_j(\bV_i)
\end{align*}
Assume $\EXP \{ |h_j(\bV_1)|\} < \infty$ for all $j$.

\begin{theorem}
\label{harro}
Assume that for given $Y=j$, $\bV_1,\cdots,\bV_t$ are conditionally independent and identically distributed. Set
\begin{align*}
q_{j,\ell} := \EXP \{ h_{\ell}(\bV_1) \mid Y=j \} \quad j,\ell \in \{1,\cdots,M\}.
\end{align*}
Suppose that
\begin{align*}
\delta := \min_{\ell \neq j }(q_{j,j} - q_{j,\ell})>0.
\end{align*}
\textbf{a)} If $ | h_j | \leq K$, $j=1,\cdots,M$, then
\begin{align*}
- \frac{1}{t} \log \PROB \{ \widehat g (\bV_1,\cdots,\bV_t )  \neq Y \} \geq  \frac{\delta^2}{2K^2}  - \frac{1}{t} \log (M-1)
\end{align*}
\textbf{b)} If $ \EXP \{ | h_\ell (\bV_1)|^k \mid Y=j \} \leq c_1^k k!$, with some $0<c_1<\infty$ for all $\ell,j=1,\cdots,M$ and all $k \in \mathbf{N}$,
then
\begin{align*}
- \frac{1}{t} \log \PROB \{ \widehat g (\bV_1,\cdots,\bV_t )  \neq Y \}
&\geq
-\log \left(1-\frac{\left(1-2\sqrt{\frac{c_1}{4c_1+\delta} } \right)^2}{2\frac{c_1}{4c_1+\delta}} \right) - \frac{1}{t} \log (M-1)\\
&\geq
\frac{\left(1-2\sqrt{\frac{c_1}{4c_1+\delta} } \right)^2}{2\frac{c_1}{4c_1+\delta}}  - \frac{1}{t} \log (M-1) \\
&\geq \frac{\delta^2}{40c_1^2} - \frac{1}{t} \log (M-1).
\end{align*}
\end{theorem}

\begin{proof}
\textbf{a)} One has that
\begin{align*}
\PROB \{ \widehat g (\bV_1,\cdots,\bV_t )  \neq Y \}
&\leq \max_j  \PROB \{ \widehat g (\bV_1,\cdots,\bV_t )  \neq j \mid Y =j \}  \\
& = \max_j  \sum_{ \ell \neq j} \PROB \{ \widehat g (\bV_1,\cdots,\bV_t )  = \ell \mid Y =j \}.
\end{align*}
For $\ell \neq j$, by Hoeffding's inequality one obtains
\begin{align}
& \PROB \{ \widehat g (\bV_1,\cdots,\bV_t )  = \ell \mid Y =j \}  \nonumber \\
&\leq  \PROB \{ \sum_{i=1}^t  h_\ell (\bV_i)  \geq \sum_{i=1}^t  h_j (\bV_i)  \mid Y =j \}  \nonumber \\
&= \PROB \{ \frac{1}{t} \sum_{i=1}^t \left [ ( h_\ell (\bV_i)- h_j (\bV_i)) - \EXP \{  ( h_\ell (\bV_i)- h_j (\bV_i))  \mid Y =j \} \right ] \geq q_{j,j}-q_{j,\ell} \mid Y=j \} \nonumber  \\
& \leq e^{ - \frac{t(q_{j,j} - q_{j,\ell})^2}{2K^2}}. \label{eq_star}
\end{align}
Then
\begin{align*}
\PROB \{ \widehat g (\bV_1,\cdots,\bV_t )  \neq Y \}
&\leq \max_{j} \sum_{\ell \neq j} e^{ - \frac{t(q_{j,j} - q_{j,\ell})^2}{2K^2}}
\leq  (M-1) \max_{\ell \neq j} e^{ - \frac{t(q_{j,j} - q_{j,\ell})^2}{2K^2}}
\end{align*}
and
\begin{align*}
- \frac{1}{t} \log \PROB \{ \widehat g (\bV_1,\cdots,\bV_t )  \neq Y \} \geq  \frac{\delta^2}{2K^2}  - \frac{1}{t} \log (M-1).
\end{align*}
\textbf{b)}
Let $\ell \neq j$.
For arbitrary $\beta>0$, one obtains
\begin{align*}
\PROB \left\{ \widehat g (\bV_1,\cdots,\bV_t )  = \ell \mid Y =j \right\}
&\leq
\PROB \left\{ \sum_{i=1}^t  (h_\ell (\bV_i)  -  h_j (\bV_i))\ge 0  \mid Y =j \right\}   \\
&\le
\EXP \left\{ e^{\beta\sum_{i=1}^t  (h_\ell (\bV_i)  -  h_j (\bV_i))}  \mid Y =j \right\}   \\
&=
\left(\EXP \left\{ e^{\beta  (h_\ell (\bV_1)  -  h_j (\bV_1))}  \mid Y =j \right\}\right)^t.
\end{align*}
Noticing
\begin{align*}
\EXP \left\{  h_\ell (\bV_1)  -  h_j (\bV_1) \mid Y =j \right\}
&=
 q_{j,\ell}-q_{j,j}
\le -\delta
<0,
\end{align*}
one has
\begin{align*}
&\EXP \left\{ e^{\beta  (h_\ell (\bV_1)  -  h_j (\bV_1))}  \mid Y =j \right\}\\
&\le
1+\beta \EXP \left\{  h_\ell (\bV_1)  -  h_j (\bV_1) \mid Y =j \right\}+\sum_{k=2}^{\infty}\frac{\beta^k}{k!}\EXP \left\{ | h_\ell (\bV_1)  -  h_j (\bV_1)|^k \mid Y =j \right\}\\
&\le
1-\beta\delta +\sum_{k=2}^{\infty}\frac{\beta^k}{k!}2^k\EXP \left\{ | h_\ell (\bV_1)|^k  +  |h_j (\bV_1)|^k \mid Y =j \right\}\\
&\le
1-\beta\delta +2\sum_{k=2}^{\infty}(2c_1 \beta)^k\\
&=
1-\beta\delta +8c_1^2 \beta^2 \frac{1}{1-2c_1 \beta}.
\end{align*}
The right-hand side is minimized for
\begin{align*}
\beta
&=
\frac{1}{2c_1} \left(1-2\sqrt{\frac{c_1}{4c_1+\delta} }\right),
\end{align*}
and the minimum value is
\begin{align*}
1-\frac{\left(1-2\sqrt{\frac{c_1}{4c_1+\delta} } \right)^2}{2\frac{c_1}{4c_1+\delta}}
\end{align*}
which exceeds $\frac{\delta^2}{40c_1^2}$ because $ 0 < \delta \leq 2c_1$.
\end{proof}

In the sequel, we apply these bounds, when the underlying conditional distributions or densities are unknown.
For the sake of illustration, we compare the error exponents, when the conditional densities $f_j$ exist and they are known.
In the case of ML decision functions, i.e. $h_ j =  \log f_j$, the term
\begin{align*}
q_{j,j}-q_{j,\ell} &= \EXP \{ h_j(\bV_1)-h_\ell(\bV_1) \mid Y=j\}
= \int \log \left( \frac{f_j(x)}{f_\ell(x)} \right) f_j(x) dx
\end{align*}
is the Kullback-Leibler divergence of the distributions  belonging to $f_j$ and $f_\ell$. As is well known,
\begin{align*}
q_{j,j}-q_{j,\ell} \geq B (f_j,f_\ell)
\end{align*}
by Jensen's inequality. The boundedness assumption  in Theorem \ref{harro}a is fulfilled if the densities $f_j$ have a common support and are there bounded away from zero and infinity. In this case neither the above inequality in context of Bhattacharyya distance nor the inequality in
Theorem \ref{harro}a is generally sharper than the other.
The moment condition in Theorem \ref{harro}b is fulfilled in the case of $d$-dimensional Gaussian distributions (with possibly different covariance matrices).

For an elementary classifier the tailor-cut proof of Theorem 1 yields an exponential rate with a constant better than $- \log (1-\min_{j \neq \ell}(\sqrt{p_{j,j}}-\sqrt{p_{j,\ell}})^2) \geq \min_{j \neq \ell} (\sqrt{p_{j,j}}-\sqrt{p_{j,\ell}})^2$, where Theorem \ref{harro}a with $h_j(\bV_i) = \IND_{g(\bV_i)=j}$ and $q_{j,\ell}=p_{j,\ell}$ yields the weaker constant
\begin{align*}
\frac{1}{2} \min_{j \neq \ell}  (p_{j,j}-p_{j,\ell} )^2,
\end{align*}
in both cases up to a common term $O\left(\frac{1}{t}\right)$.

\begin{prop}
\label{CLT2}
Assume the conditions of Theorem \ref{harro}. If $ \EXP \{  h_j(\bV_1)^2 \} <\infty$ for all $j=1,\cdots,M$,  then the CLT implies that
\begin{align*}
\PROB\{\widehat g(\bV_1,\dots ,\bV_t)=\ell\mid Y=j\}
&\le
\PROB\Big\{\sum_{i=1}^t \left(h_\ell (\bV_i)-h_j (\bV_i)\right)\ge 0\mid Y=j\Big\}\\
&\approx
\Phi\left(-\frac{\sqrt{t}(q_{j,j}-q_{j,\ell})}{\sqrt{\Var\left(h_\ell (\bV_1)-h_j (\bV_1)\mid Y=j\right)}}\right).
\end{align*}
If, in addition, $ \EXP \{ | h_j(\bV_1)|^3 \}<\infty$ for all $j=1,\cdots,M$, then because of the Berry-Esseen inequality, the approximation error is of order $O(1/\sqrt{t})$.
\end{prop}

\section{Classification by nominal densities.}
\label{Nomi}

Devroye, Gy\"orfi and Lugosi \cite{DeGyLu02} introduced  a robust detection rule, which may result in a classification algorithm for multiple classes.
Let $f^{(1)},\ldots,f^{(M)}$ be fixed densities with respect to a dominating measure $\lambda$ on $\Rd$, which
are the nominal densities under $M$  classes.
We observe i.i.d. random vectors $\bV_1,\dots ,\bV_t$ according
to a common density $f$. Under the class $j$,
 the density $f$ is a
distorted version of the nominal density $f^{(j)}$ for class $j$, which means that  there exists a positive
number $\epsilon$ such that
\begin{align}
\label{ep}
  \|f-f^{(j)}\| \le \Delta_j - \epsilon,
\end{align}
where $\Delta_j := (1/2)\min_{i\neq j} \|f^{(i)}-f^{(j)}\|$.
Here $\|f-g\| = \int |f-g|d\lambda$ denotes the $L_1$ distance
between two densities.
Introduce the empirical measure
\[
  \mu_t(A) = \frac{1}{t} \sum_{i=1}^t \IND_{\bV_i\in A}~,
\]
where  $A$ is
a Borel set.
Let $\A$ denote the collection of $M(M-1)/2$ sets
of the form
\[
  A_{i,j}= \left\{\bx:f^{(i)}(\bx)>f^{(j)}(\bx)\right\}~, \quad 1\le i<j \le M~.
\]
The corresponding classification rule is as follows:
\begin{align}
\label{tl}
\tilde g(\bV_1,\dots ,\bV_t)
=j \mbox{ if } \max_{A\in\A} \left|\int_A f^{(j)}d\lambda -\mu_t(A)\right|
  =  \min_{i=1,\ldots,M} \max_{A\in\A} \left|\int_A f^{(i)}d\lambda -\mu_t(A)\right|~.
\end{align}
Devroye, Gy\"orfi and Lugosi \cite{DeGyLu02} proved that,
for any $f$ satisfying (\ref{ep}),
\begin{align}
\label{DGL}
\PROB\{ \tilde g(\bV_1,\dots ,\bV_t)\ne Y\mid Y=j\}
\le  2M(M-1)^2 e^{-t\epsilon ^2/2}.
\end{align}

The classifier $\hat g$, defined by (\ref{tl}) supposes the calculation of $\int_A f^{(j)}d\lambda$.
For $M=2$, Gy\"orfi and Walk \cite{GyWa14} presented a simplified robust detection rule, which does not need such integrals.
In fact, their rule is a maximum likelihood classifier with respect to the nominal densities.

Motivated by robust detection algorithms studied in Devroye, Gy\"orfi and Lugosi \cite{DeGyLu02}
 and in Biglieri and Gy\"orfi \cite{BiGy14}, Afser \cite{Afs2021} introduced a classification rule similar to $\tilde g$.
If $\bX$ takes finitely many values, then the nominal discrete distributions are replaced by distribution estimates derived from $\D_n$ and he gave exponential upper bound on the error probability $\PROB\{\tilde g(\bV_1,\dots ,\bV_t)\ne Y\}$.

Based on these nominal densities, we consider two classifiers.
Put
\begin{align}
\label{fj}
A_j
&=
\{\bx: f^{(j)}(\bx)>f^{(\ell)}(\bx), \ell\ne j\}.
\end{align}
The {\em aggregated nominal maximum likelihood classifier} is defined by
\begin{align*}
\tilde g(\bV_1,\dots ,\bV_t)
&=
\argmax_j \sum_{i=1}^t \IND_{\bV_i\in A_j},
\end{align*}
For the notation
\begin{align*}
p_{j,\ell}=\PROB\{\bV_1\in A_\ell\mid Y=j\},
\end{align*}
the conditional error probabilities can be bounded as in Proposition \ref{CLT1}:
\begin{align}
\label{clt1}
\PROB\{\tilde g(\bV_1,\dots ,\bV_t)=\ell\mid Y=j\}
&\lessapprox
\exp\left(-\frac{t(p_{j,j}-p_{j,\ell})^2}{2\Var (\IND_{\bV_1\in A_{\ell}}- \IND_{\bV_1\in A_j}\mid Y=j)}\right).
\end{align}
If $f^{(j)}(\bx)>0$ for all $j$ and $\bx$, then the {\em nominal log-maximum likelihood classifier} is
\begin{align*}
\widehat g (\bV_1,\cdots,\bV_t ) = \argmax_{j} \sum_{i=1}^t \log(f^{(j)}(\bV_i))
\end{align*}
and its conditional error probabilities follow from
Proposition \ref{CLT2}:
\begin{align*}
\PROB\{\widehat g(\bV_1,\dots ,\bV_t)=\ell\mid Y=j\}
&\lessapprox
\exp\left(-\frac{t(q_{j,j}-q_{j,\ell})^2}{2\Var\left(\log(f^{(\ell)} (\bV_1))- \log(f^{(j)}(\bV_1))\mid Y=j \right)}\right),
\end{align*}
where $q_{j,\ell} := \EXP \{ \log(f^{(\ell)} (\bV_1)) \mid Y=j \}$.

\section{Prototype classification.}
\label{Proto}

If the nominal densities are shifted versions of standard normal density, then the resulted classifier is a particular prototype classifier defined as follows:
for the prototype vectors $\bx_1,\dots ,\bx_M\in \R^d$,
introduce the prototype classifier as
\begin{align*}
\widehat g_{pr} (\bV_1,\cdots,\bV_t ) = \argmin_{j} \sum_{i=1}^t \|\bV_i-\bx_j\|^2.
\end{align*}
For the notation
\begin{align*}
h_j(\bx)=-\|\bx-\bx_j\|^2,
\end{align*}
the conditional error probabilities can be calculated by Proposition \ref{CLT2}:
\begin{align}
\label{cltpr}
\PROB\{\widehat g_{pr}(\bV_1,\dots ,\bV_t)=\ell\mid Y=j\}
&\lessapprox
\exp\left(-\frac{t\EXP\{\|\bX-\bx_{\ell}\|^2-\|\bX-\bx_j\|^2\mid Y=j\}^2}{2\Var\left(\|\bX-\bx_{\ell}\|^2-\|\bX-\bx_j\|^2\mid Y=j \right)}\right).
\end{align}

If the prototype vectors $\bx_1,\dots ,\bx_M$ are the class-conditional expectations
\begin{align*}
\bx_j
&=\EXP \{ \bX \mid Y=j \},
\end{align*}
then
\begin{align*}
\EXP\{\|\bX-\bx_{\ell}\|^2-\|\bX-\bx_j\|^2\mid Y=j\}
&=
2\EXP\{(\bX,\bx_j-\bx_{\ell})\mid Y=j\}+\|\bx_{\ell}\|^2-\|\bx_j\|^2\\
&=
\|\bx_j-\bx_{\ell}\|^2
\end{align*}
and
\begin{align*}
\Var\left(\|\bX-\bx_{\ell}\|^2-\|\bX-\bx_j\|^2\mid Y=j \right)
&=
4\Var\left( (\bX,\bx_j-\bx_{\ell})\mid Y=j \right)\\
&=
4(\bx_j-\bx_{\ell}, \bC_j(\bx_j-\bx_{\ell})),
\end{align*}
with the $j$-th conditional covariance matrix $\bC_j$ of $\bX$.

Thus, from (\ref{cltpr}) one gets
\begin{align}
\label{cltC}
\PROB\{\widehat g_{pr}(\bV_1,\dots ,\bV_t)=\ell\mid Y=j\}
&\lessapprox
\exp\left(-\frac{t\|\bx_j-\bx_{\ell}\|^4}{8(\bx_j-\bx_{\ell}, \bC_j(\bx_j-\bx_{\ell}))}\right) \nonumber\\
&\le  \exp\left(-\frac{   t  \|\bx_j-\bx_{\ell}\|^2 }{8  \text{tr} \left( \bC_j \right)}\right),
\end{align}
where the last inequality follows from the relations
\begin{align*}
\left( \bx_j - \bx_\ell, \bC_j(\bx_j - \bx_\ell) \right) &= \text{tr} \left( \bC_j(\bx_j - \bx_\ell)(\bx_j - \bx_\ell)^\t \right) \\
& \leq \text{tr}(\bC_j) \text{tr}  \left((\bx_j - \bx_\ell)(\bx_j - \bx_\ell)^\t \right) \\
&=  \text{tr}(\bC_j) \|\bx_j-\bx_{\ell}\|^2
\end{align*}
for traces of matrices and their product.

One can improve the performance of the prototype classifier by appropriate  transformation of the feature vector.
Introduce a linear transformation $T$ of $\R^d$ to $\R^M$ such that
\begin{align*}
T(\bx)&=\A^\t \bx,
\end{align*}
where $\A \in \R^{d \times M}$ is the transformation matrix with columns $\ba_j \in \R^d$, $j=1,\dots,M$.
The vectors $\ba_j$ can be interpreted as scaling vectors.
Put
\[
\widehat g_{T}(\A^\t \bV_1,\dots ,\A^\t \bV_t)
= \argmax_j \sum_{i=1}^t \|\A^\t (\bV_i-\bx_j)\|^2.
\]
As before,
\begin{align*}
\EXP\{\|\A^\t (\bX-\bx_{\ell})\|^2-\|\A^\t (\bX-\bx_j)\|^2\mid Y=j\}
&=
\|\A^\t (\bx_j-\bx_{\ell})\|^2
\end{align*}
and
\begin{align*}
\Var\left(\|\A^\t (\bX-\bx_{\ell})\|^2-\|\A^\t (\bX-\bx_j)\|^2\mid Y=j \right)
&=
4\Var\left( (\A^\t \bX,\A^\t (\bx_j-\bx_{\ell}))\mid Y=j \right)\\
&=
4\left( \A^\t (\bx_j-\bx_{\ell}), (\A^\t \bC_j \A) \A^\t (\bx_j-\bx_\ell) \right)
\end{align*}
noticing that the $j$-th conditional covariance matrix of $\A^\t \bX$ is $ \A^\t \bC_j \A$.

Correspondingly to (\ref{cltC}), we obtain
\begin{align}
\label{cltA}
\PROB\{\widehat g_{T}(\A^\t \bV_1,\dots ,\A^\t \bV_t)=\ell\mid Y=j\}
&\lessapprox
\exp \left(-\frac{t \| \A^\t (\bx_j - \bx_\ell \|^4 }{8 \left( \A^\t (\bx_j-\bx_{\ell}), (\A^\t \bC_j \A) \A^\t (\bx_j-\bx_\ell) \right)  }\right) \nonumber\\
&\le
\exp\left(-\frac{t \| \A^\t (\bx_j - \bx_\ell) \|^2 }{8 \text{tr}( \A^\t \bC_j \A )  }\right),
\end{align}
where the last inequality results from
\begin{align*}
\left( \A^\t (\bx_j-\bx_{\ell}), (\A^\t \bC_j \A) \A^\t (\bx_j-\bx_\ell) \right)&= \text{tr} \left(  (\A^\t \bC_j \A) \A^\t (\bx_j-\bx_\ell)(\bx_j-\bx_{\ell})^\t \A \right) \\
& \le \text{tr} ( \A^\t \bC_j \A ) \text{tr} \left( \A^\t (\bx_j-\bx_\ell)(\bx_j-\bx_{\ell})^\t \A \right) \\
&=  \text{tr} (  \A^\t \bC_j \A ) \| \A^\t (\bx_j - \bx_\ell \|^2.
\end{align*}
Therefore
\begin{align*}
\PROB\{\widehat g_{T}(\A^\t \bV_1,\dots ,\A^\t \bV_t) \neq Y\} &\leq \max_{j} \PROB\{\widehat g_{T}(\A^\t \bV_1,\dots ,\A^\t \bV_t) \neq j \mid Y = j\}  \\
& =  (M-1)  \max_{j} \max_{\ell \neq j} \PROB\{\widehat g_{T}(\A^\t \bV_1,\dots ,\A^\t \bV_t) = \ell \mid Y = j\}   \\
& \lessapprox
(M-1)     \max_{\ell \neq j}  \exp\left(-\frac{t \| \A^\t (\bx_j - \bx_\ell) \|^2 }{8 \text{tr}( \A^\t \bC_j \A )  }\right)\\
& =
(M-1)      \exp\left(-\frac{t  }{8 \sigma^2(\A)  }\right)
\end{align*}
with
\begin{align*}
\sigma^2(\A)
&=
\max_{\ell \neq j}  \frac{ \text{tr}( \A^\t \bC_j \A )  }{ \| \A^\t (\bx_j - \bx_\ell) \|^2 } .
\end{align*}

The criterion $\sigma^2(\A)$ is hard to minimize, therefore
we consider two criteria, which are related to $\sigma^2(\A)$ via $\text{tr}( \A^\t \bC_j \A )=\EXP\{\|\A^\t (\bX- \bx_j)\|^2\mid Y=j\}$.
Firstly, put
\begin{align*}
\sigma_1^2(\A)
&=
\frac{\sum_{j=1}^{M}\EXP\{\|\A^\t (\bX- \bx_j)\|^2\mid Y=j\}}{\sum_{j=1}^{M}\|\A^\t (\bar \bx- \bx_j)\|^2 },
\end{align*}
where $\bar \bx = \frac{1}{M}\sum_{k=1}^M \bx_k$.
Secondly,
choose the matrix $\A$ as follows:
for an arbitrary orthonormal set of $M$-dimensional vectors $\be_1,\be_2,\dots,\be_M$, set
$\A^\t \bx_j=\be_j$.
Then, $\|\A^\t \bx_{\ell}- \A^\t \bx_j\|^2=\|\be_{\ell}- \be_j\|^2=2$ and therefore
 the second error criterion is defined by
\begin{align*}
\sigma_2^2(\A)
&=
\sum_{j=1}^{M}\EXP\{\|\A^\t (\bX- \bx_j)\|^2\mid Y=j\}.
\end{align*}
In the sequel, we develop algorithms for minimizing these criteria.
If $\A_i=\argmin_{\A}\sigma_i(\A)$, $i=1,2$, then one may select the matrix $\A$ out of $\A_1$ and $\A_2$,
which yields $\min\{\sigma(\A_1),\sigma(\A_2)\}$.

\textcolor{black}{Concerning $\sigma_1$, one has that
\begin{align*}
\EXP\{\|\A^\t (\bX- \bx_j)\|^2\mid Y=j\}
&=
\EXP\{\|\A^\t (\bX- \EXP\{\bX\mid Y=j\})\|^2\mid Y=j\}\\
&=
\sum_{i=1}^M\EXP\{(\ba_i,\bX- \EXP\{\bX\mid Y=j\})^2\mid Y=j\}\\
&=
\sum_{i=1}^M(\ba_i,\bC_j\ba_i)
\end{align*}
with the $j$-th conditional covariance matrix $\bC_j$, and
\begin{align*}
\|\A^\t (\bar \bx-\bx_j )\|^2
=
\sum_{i=1}^M(\ba_i,\bar \bx-\bx_j)^2.
\end{align*}
Thus,
\begin{align}
\label{fact}
\sigma_1^2(\A)
&=
\frac{\sum_{j=1}^M\sum_{i=1}^M(\ba_i,\bC_j\ba_i)  }{\sum_{j=1}^M \sum_{i=1}^M(\ba_i,\bar \bx-\bx_j)^2 }
=
\frac{ \sum_{i=1}^M(\ba_i, \bS_W\ba_i)  }{ \sum_{i=1}^M(\ba_i,\bS_B\ba_i) },
\end{align}
where
\begin{align*}
\bS_W &= \sum_{j=1}^M \bC_j,   \\
\bS_B  &=  \sum_{j=1}^M  (\bar \bx- \bx_j)(\bar \bx- \bx_j)^\t.
\end{align*}}
For the $d \times M$ matrix $\A$, we have
\begin{align*}
\A^\t S_W \A
&= \begin{pmatrix} \ba_1^\t \\ \ba_2^\t \\ \vdots \\ \ba^\t_M \end{pmatrix}
\bS_W \left( \begin{array}{cccc}
\ba_1 & \ba_2 & \hdots & \ba_M \\
\vdots  & \vdots &  & \vdots
\end{array} \right )\\
&= \left( \begin{array}{cccc}
( \ba_1,\bS_W \ba_1) & ( \ba_1,\bS_W \ba_2) & \hdots &  ( \ba_1,\bS_W \ba_M) \\
( \ba_2,\bS_W \ba_1) & ( \ba_2,\bS_W \ba_2) &   \hdots & ( \ba_2,\bS_W \ba_M) \\
\vdots & \vdots & \ddots & \vdots \\
( \ba_M,\bS_W \ba_1) & ( \ba_M,\bS_W \ba_2)&   \hdots & ( \ba_M,\bS_W \ba_M) \
\end{array} \right )
\end{align*}
Thus, $\A^\t S_W \A $ is an $M \times M$ matrix and the sum of its diagonal terms equals
\begin{align}
\label{13*}
\text{tr}( \A^\t S_W \A ) = \sum_{i=1}^M (\ba_i, \bS_W \ba_i)
\end{align}
and similarly
\begin{align*}
 \text{tr}( \A^\t S_B \A ) = \sum_{i=1}^M (\ba_i, \bS_B \ba_i).
\end{align*}
Therefore, by (\ref{fact}),
\begin{align}
\label{13**}
\sigma_1^2(\A)=\frac{ \text{tr}( \A^\t \bS_W \A) }{ \text{tr} ( \A^\t \bS_B \A )  }.
\end{align}
\textcolor{black}{Next, we consider the optimization of the right hand side of (\ref{13**}).
Put
\begin{align*}
\rho(\A)=\frac{ \text{tr} ( \A^\t \bS_B \A )  }{ \text{tr}( \A^\t \bS_W \A) }
\end{align*}
and
\begin{align*}
\rho^*=\max_{\A}\rho(\A).
\end{align*}
$\rho^*$ is unique, however, the optimal matrix $\A$ is not unique.
Set
\begin{align*}
D=\{\A:\rho(\A)=\rho^*\}.
\end{align*}
(\ref{fact}) implies that if $\A\in D$, then any permutation of its columns is a solution, too.
Therefore, the set $D$ is a union of at least $M!$ disconnected sets.}

In the sequel, we present an optimization algorithm under the constraint $\A^\t \A = \mathbb{I}_M $
The existence of the maximizer $\A$
implies $ \A^\t (\bS_B -\rho^*\bS_W) \A=0$. Consequently, the optimal $\A$ consists of $M$ eigenvectors of $(\bS_B-\rho^*\bS_W)$ with largest magnitude eigenvalues whose sum must be equal to zero,
see Ngo,  Bellalij and Saad \cite{Ngo12}.

Here is the algorithm:\\
Select an  arbitrary initial matrix $\A_0$ with $\A_0^\t \A_0 = \mathbb{I}_M$.
Put  $\rho_0= \frac{ \text{tr} ( \A_0^\t \bS_B \A_0 )  }{ \text{tr}( \A_0^\t \bS_W \A_0) }$.\\
For $k=1,2,\dots$, introduce the iteration:\\
Compute the $M$ largest eigenvalues of $\bS_B-\rho_{k-1}\bS_W$ and set $\ba_{k,1},\dots,\ba_{k,M}$ to be the corresponding eigenvectors. $\A_k$ consists of these eigenvectors.\\
Set
\begin{align*}
\rho_k=\frac{ \text{tr} ( \A_k^\t \bS_B \A_k )  }{ \text{tr}( \A_k^\t \bS_W \A_k) }.
\end{align*}

\textcolor{black}{\begin{prop}[Theorem 5.1 in Zhang et al. \cite{ZhLiNg10}]
The sequence $\rho_k$ generated by the above algorithm is monotonically increasing to $\rho^*$ such that
\begin{align*}
0
\leq \rho^* - \rho_{k+1}
\leq (1-\gamma)(\rho^* - \rho_k)
\leq (1-\gamma)^k(\rho^* - \rho_0),
\end{align*}
where
\begin{align*}
\gamma =  \frac{\sum_{i=1}^M \lambda_{d-i+1}(\bS_W)}{\sum_{i=1}^M \lambda_{i}(\bS_W)} \in (0,1]
\end{align*}
and $\lambda_i ( \cdot)$ denotes the $i$-th largest eigenvalue.
\end{prop}}
\textcolor{black}{This proposition is on the convergence of $\rho(\A_k)=\frac{ \text{tr} ( \A_k^\t \bS_B \A_k )  }{ \text{tr}( \A_k^\t \bS_W \A_k) }$.
However, the sequence $\A_k$ is not necessarily convergent.
Note that we do not need the convergence of $\A_k$.}

As to $\sigma_2$, correspondingly to (\ref{fact}), we have that
\begin{align}
\label{13***}
\sigma_2^2(\A)
&=
\sum_{i=1}^M(\ba_i,\bS_W\ba_i).
\end{align}
Set
\begin{align*}
\bQ^\t =
\begin{pmatrix}
\bx_1^\t \\
\bx_2^\t \\
\vdots \\
\bx_M^\t
\end{pmatrix},
\quad
\bE^\t =
\begin{pmatrix}
\be_1^\t \\
\be_2^\t \\
\vdots \\
\be_M^\t
\end{pmatrix}
\end{align*}
The system of equations $\A^\t \bx_1 = \be_1, \dots,\A^\t \bx_M = \be_M$, can be expressed as
\begin{align*}
\A^\t  \left( \begin{array}{cccc}
\bx_1 & \bx_2 & \hdots & \bx_M \\
\vdots  & \vdots &  & \vdots
\end{array} \right ) &=  \left( \begin{array}{cccc}
\be_1 & \be_2 & \hdots & \be_M \\
\vdots  & \vdots &  & \vdots
\end{array} \right ), 
\end{align*}
i.e.,
\begin{align*}
\A^\t \bQ&= \bE,
\end{align*}
where $ \bE$ is a unitary matrix consisting of  orthonormal columns $\be_1,\be_2,\dots,\be_M$, so that  $\bE^\t \bE = \mathbb{I}_M$. But,  since $\bE$ is square the rows are orthonormal as well, i.e., $\bE \bE^\t=  \mathbb{I}_M$. Thus
\begin{align*}
\A^\t \bQ \bQ^\t \A = \bE \bE^\t = \mathbb{I}_M.
\end{align*}
Let
\begin{align*}
 \bS_C   = \bQ \bQ^\t =  \left( \begin{array}{cccc}
\bx_1 & \bx_2 & \hdots & \bx_M \\
\vdots  & \vdots &  & \vdots
\end{array} \right ) \begin{pmatrix}
\bx_1^\t \\
\bx_2^\t \\
\vdots \\
\bx_M^\t
\end{pmatrix} = \sum_{i=1}^M \bx_i \bx_i^\t,
\end{align*}
so that $ \bS_C$ is a $d \times d $ matrix.

Then, noticing also (\ref{13*}) and (\ref{13***}), compare (\ref{13**}),
the optimization problem is
 \begin{align*}
\min_{\A, \A^\t  \bS_C  \A = \mathbb{I}_M}  \sigma_2^2(\A) = \min_{\A, \A^\t  \bS_C  \A = \mathbb{I}_M}   \text{tr} ( \A^\t \bS_W \A )
\end{align*}
The exact solution to this problem is known.
It concerns the generalized eigenvalue problem and $\A$ consists of the eigenvectors of $\bS_W^{-1}\bS_C$ with largest eigenvalues.

The Lagrange function $\varphi$ for the optimization problem is
\begin{align*}
\varphi (\A, \bL) = \text{tr}(\A^\t \bS_W \A)  + \bL(\mathbb{I}_M - \A^\t  \bS_C  \A)
\end{align*}
where $\bL=\text{diag}(\l_1,\cdots,\l_M)$ is the coefficient matrix. In order to have stationary points, the first order conditions require
\begin{align}
\frac{\partial \varphi (\A, \bL)}{\partial \ba_i} = 0, \quad  \quad \frac{\partial \varphi (\A, \Lambda)}{\partial \l_i} =0, \quad  i=1,\dots ,M . \label{nec_conds}
\end{align}
The first set of conditions above yields $(\bS_W  - \l_i \bS_C)^\t\ba_i=0$, $i=1,\dots ,M$ and is equivalent to the generalized eigenvalue problem $(\bS_C- \frac{1}{\l_i}\bS_W)^\t \ba_i=0$, 
since $\bS_W$ has full rank, and one only needs to consider the case $\l_i \neq 0$ in order to have a solution other than $\ba_i=[0,\cdots,0]$. Thus, $(\ba_i,\frac{1}{\l_i})$, $\l_i \neq 0$, are the (generalized) eigenvector/eigenvalue pairs of $(\bS_C, \bS_W)$ or equivalently $\bS_W^{-1}\bS_C$, since $\bS_W$ is invertible. Simultaneously diagonalization procedure \cite[Ch. 2]{Fuk90} implies that there exists a non-singular matrix $\bB$  such that
\begin{align*}
\bB^\t \bS_W \bB &= \mathbb{I}_d, \\
\bB^\t  \bS_C \bB &= \Lambda = \text{diag}(\lambda_1,\cdots,\lambda_M,0,\cdots,0).
\end{align*}
Moreover, the columns of $\bB$, $\bb_1,\cdots,\bb_d$, are the eigenvectors of $(\bS_C, \bS_W)$  with corresponding eigenvalues in $\Lambda$. 
Therefore, the first set of condition in \eqref{nec_conds} is satisfied when $(\ba_i,\frac{1}{\l_i})=(\bb_i,\lambda_i)$, $i=1,\cdots,M$, while second set of condition in \eqref{nec_conds} yields
\begin{align*}
\ba_i^\t  \bS_C \ba_i=1
\end{align*}
and gives the necessary scaling on $\ba_i$. Simultaneous diagonalization indicates  $\bb_i^\t \bS_C \bb_i = \lambda_i$, $i=1,\cdots,M$, thus we conclude that both sets of the first order conditions in \eqref{nec_conds} are satisfied  when
\begin{align}
 \label{sol_sigma1}
\ba_i = \frac{ \bb_i}{\sqrt{\lambda_i}}, \quad i=1,\cdots,M,
\end{align}
and results in
\begin{align*}
\min_{\A, \A^\t  \bS_C  \A = \mathbb{I}_M}  \sigma_2^2(\A) = \min_{\A, \A^\t  \bS_C  \A = \mathbb{I}_M}   \text{tr} ( \A^\t \bS_W \A ) = \sum_{i=1}^M \frac{1}{\lambda_i}.
\end{align*}

\section{Linear classification}
\label{Class}

In this section let $M=2$.
For the sake of simplicity, assume that $Y$ is $\pm 1$ valued.
Put
\[
D(\bx)=\EXP\{Y \mid \bX=\bx\}.
\]
Then,
the Bayes decision $g^*$ is of the form
\[
g^*(\bx) =\mbox{sign }D(\bx).
\]

For $M=2$, the prototype classifier is a linear classifier.
If the features are different physical quantities, then this elementary linear classifier may have poor performance.
Therefore, there is a space for improving the linear classifier.

The  linear classification, called linear discrimination, too, has a long history, see for example Chapter 4 in Devroye, Gy\"orfi and Lugosi
\cite{DeGyLu96}.
It started with the classical concept of perceptron due to Rosenblatt \cite{Ros62}.
Linear discrimination is at the heart of nearly every successful pattern recognition
method, including tree classifiers, generalized linear classifiers, neural networks, etc.

For a weight vector $\bc=(c_0,c_1,\dots ,c_d)$ and for the observation vector $\bX=(X_1,\dots ,X_d)$, the corresponding linear classifier is defined by
\[
g_{\bc}(\bX) =\mbox{sign}\left(c_0+ \sum_{i=1}^d c_iX_i\right).
\]
The obvious aim here is to achieve
\[
P_{e,lin}^*=\PROB\{g_{\bc^*}(\bX)\ne Y\}=\min_{\bc} P_e(g_{\bc}).
\]
In general, $P_e(g_{\bc})$ is not a unimodal function of $\bc$.
Interestingly, if $P_{e}^*=P_{e,lin}^*$, then $P_e(g_{\bc})$ is unimodal
such that $P_e(g_{\bc})$ is monotone increasing along rays pointing from $\bc^*$,
cf. Fritz and Gy\"orfi \cite{FrGy75}.

Next, we study the simple algorithms for getting linear classifiers with good performance.
Instead of searching for a separating hyperplane given by the weight vector $\bc=(c_0,c_1,\dots ,c_d)$, we fix the  hyperplane corresponding to the particular weight vector $\bc=(0,1,\dots ,1)$ and construct a rescaling of the observation vector, for which the error probability is small.
Introduce a rescaling of the observation vector $\bX=(X_1,\dots ,X_d)$
such that under the event $\{Y=-1\}$ the transformed vector is concentrated around
$(-1,-1,\dots ,-1)$, while under $\{Y=1\}$ it is around $(1,1,\dots ,1)$.

Assume that the class conditional expectations of the features can be estimated from data, and so we assume that these expectations are known.
Let $\PROB_-$, $\EXP_-$ and $\PROB_+$, $\EXP_+$ denote the conditional probability distribution and the conditional expectation under the events $\{Y=-1\}$ and $\{Y=1\}$, respectively.

Let $\bZ=(Z_1,\dots ,Z_d)$ denote the rescaled vector such that
\[
Z_i=\frac{2}{m_{+,i}-m_{-,i}}(X_i-m_{+,i}) +1,
\]
where
\[
m_{-,i}=\EXP_-\{X_i\}
\mbox{\quad and \quad}
m_{+,i}=\EXP_+\{X_i\}
\]
and we assume that
\[
m_{-,i}\ne m_{+,i},
\]
$i=1,\dots ,d$.
Then,
\[
\EXP_-\{Z_i\}=-1
\mbox{\quad and \quad}
\EXP_+\{Z_i\}=1.
\]

For this rescaling and for the notation
\begin{align}
\label{Gsig}
G(\bX)
=V
=\frac 1d \sum_{i=1}^d Z_i
=\frac 2d \sum_{i=1}^d \frac{X_i-m_{+,i}}{m_{+,i}-m_{-,i}} +1,
\end{align}
we have that
\[
\EXP_-\left\{V\right\}
=-1
\mbox{\quad and \quad}
\EXP_+\left\{V\right\}
=1.
\]

Introduce the linear classifier
\begin{align}
\label{sig}
g(\bX) =\mbox{sign}\left( V\right),
\end{align}
and using the notation (\ref{Gsig}) the corresponding aggregated classifier is defined by
\begin{align}
\label{Gt}
\widehat g(\bV_1,\dots ,\bV_t)
&= \mbox{sign}\left( \sum_{i=1}^t G(\bV_i) \right).
\end{align}

Set
\begin{align*}
\sigma^2_{d,-}=\Var_-\left(\sum_{i=1}^d Z_i/d\right)
\quad \mbox{and} \quad
\sigma^2_{d,+}=\Var_+\left(\sum_{i=1}^d Z_i/d\right).
\end{align*}
Similarly to Proposition \ref{CLT2}, we can derive a CLT approximation:
\begin{prop}
\label{lrep}
Assume that for given $Y=-1$ and $Y=1$,   $\bV_1,\dots ,\bV_t $ are conditionally independent and identically distributed, respectively.
If the conditional distributions of $V$ are approximately normal, then
\begin{align*}
\PROB\{\widehat g(\bV_1,\dots ,\bV_t) \ne Y\}
&\lessapprox
2e^{-t/(2\max\{\sigma^2_{d,-},\sigma^2_{d,+} \} )}.
\end{align*}
\end{prop}

\textcolor{black}{Under mild conditions,  $\sigma^2_{d,+}\le c/d$ and $\sigma^2_{d,-}\le c/d$.
For example, if the features are conditionally uncorrelated and the sequence of conditional variances of the features is bounded, then we have these inequalities.
Therefore, one can bound the error exponent such that $1/(2\max\{\sigma^2_{d,-},\sigma^2_{d,+} \} )\ge d/(2c)$,
from which the bound in the proposition has the following form:
\begin{align*}
\PROB\{\widehat g(\bV_1,\dots ,\bV_t) \ne Y\}
&\lessapprox
2e^{-dt/(2c )}.
\end{align*}
It means that the bound is exponentially decreasing, when $d$ is increasing.}

\bigskip
In the sequel we assume that the conditional covariance matrices are known, too.
The problem left is how to utilize these covariances for improving the previous scaling.
We introduce a linear transformation $T$ of $\R^d$ to $\R$ such that
\begin{align*}
T(\bz)=(\ba,\bz),
\end{align*}
where $\ba$ is a scaling vector, i.e., this particular transformation is just a scaling.
With the notation
\begin{align*}
\bz_+=\EXP\{\bZ\mid Y=1\}
\quad
\mbox{and}
\quad
\bz_-=\EXP\{\bZ\mid Y=-1\},
\end{align*}
for a scaling vector $\ba$, put
\begin{align*}
\sigma^2(\ba)
&:=
\frac{\max\{\EXP\{|(\ba,\bZ)- (\ba,\bz_+)|^2\mid Y=1\},\EXP\{|(\ba,\bZ)- (\ba,\bz_-)|^2\mid Y=-1\} \}}{|(\ba,\bz_{+})- (\ba,\bz_-)|^2 }.
\end{align*}

Set
\[
\widehat  g_{\ba}((\ba,\bV_1),\dots ,(\ba,\bV_t))
= \mbox{sign}\left( \sum_{i=1}^t (\ba,\bV_i)\right).
\]
Similarly to Proposition \ref{lrep}, one can show that
\begin{align}
\label{CCh}
\PROB\{\widehat g_{\ba}((\ba,\bV_1),\dots ,(\ba,\bV_t)) \ne Y\}
&\lessapprox
2e^{-t/(2\sigma(\ba)^2 )}.
\end{align}
The task left is to construct
\begin{align*}
\ba^*_d:=\argmin_{\ba} \sigma^2(\ba).
\end{align*}
If $\bC_+$ and $\bC_-$ denotes the conditional covariance matrix of $\bZ$ given $Y=1$ and $Y=-1$, respectively, then
\begin{align*}
\sigma^2(\ba)
=
\frac{\max\{(\ba,\bC_+\ba),(\ba,\bC_-\ba)\}}{(\ba,\bz_{+}-\bz_-)^2 }.
\end{align*}
Instead of $\sigma^2(\ba)$ we look at the easier quantity
\begin{align*}
\tilde\sigma^2(\ba)
=
\frac{(\ba,(\bC_+ + \bC_-)\ba)}{(\ba,\bz_{+}-\bz_-)^2 }.
\end{align*}
If $\b1$ stands for the all $1$ vector, then  \textcolor{black}{$\bz_{+}-\bz_-=2\cdot \b1$ and so
\begin{align*}
\sigma^2(\b1 )
&=
\frac{\max\{(\b1,\bC_+ \b1),(\b1, \bC_-\b1)\}}{4d^2 }
=
\max\{\sigma^2_{d,-},\sigma^2_{d,+} \}/4
\end{align*}
and likewise
\begin{align*}
\tilde\sigma^2(\b1 )
&=
(\sigma^2_{d,-}+\sigma^2_{d,+})/4.
\end{align*}}
Let $\bS= \bC_+ + \bC_-$. Then,
\begin{align*}
\tilde\sigma^2(\ba)
&=  \frac{(\ba,\bS\ba)}{(\ba,\bz_{+}-\bz_-)^2 }
= \frac{1}{J(\ba)},
\end{align*}
where $J(\ba)$ is the objective function of the Fisher Linear Discriminant Analysis (LDA). Therefore minimizing $\tilde\sigma^2(\ba)$ is equivalent to maximizing $J(\ba)$.
($J(\ba)$ is called signal-to-noise ratio, too.)
It is known that
\begin{align*}
\tilde\ba^*_d
:=\argmin_{\ba} \tilde\sigma^2(\ba)
= \bS^{-1}(\bz_+ - \bz_-)
= 2\cdot \bS^{-1}\b1,
\end{align*}
\textcolor{black}{see  p. 47 in \cite{DeGyLu96} or p. 233 in \cite{DuHaSt12}.
Then,
\begin{align*}
\tilde\sigma^2(\tilde\ba^*_d)
&=
\frac{(\bS^{-1}\b1,\b1)}{4(\bS^{-1}\b1,\b1)^2 }
=
\frac{1}{4(\bS^{-1}\b1,\b1) }.
\end{align*}
Furthermore,
\begin{align*}
\tilde\sigma^2(\b1 )
&\ge
\tilde\sigma^2(\tilde\ba^*_d)
\end{align*}
and
\begin{align*}
\tilde\sigma^2(\tilde\ba^*_d)
&\ge
\tilde\sigma^2(\tilde\ba^*_{d+1}).
\end{align*}
It means that for the exponential decrease in the bound (\ref{CCh}), the error exponent is increasing, when $d$ is increasing.}

%




\end{document}